\documentstyle[multicol,aps,pre,epsf]{revtex}

\begin{document}

\def\s{\sigma}
\def\t{\tau}
\def\l{\lambda}
\def\r{\rho}
\def\m{\mu}
\def\n{\nu}
\def\e{\epsilon}
\def\om{\omega}
\def\p{\pi}

\title
{Lattice Boltzmann method for viscoelastic fluids}

\author{Iaroslav Ispolatov$^{1,2} and $ Martin Grant$^{1}$}
\address { $^{1}$Department of Physics, McGill University,
3600 rue University, Montr\a'eal,
Qu\a'ebec, Canada, H3A 2T8\\
$^2$Center for Studies in physics and Biology, Rockefeller University,
1230 York Ave, New York, NY 10021, USA.} 
\date{\today}
\maketitle

\begin{abstract}
Lattice Boltzmann model for viscoelastic flow simulation is proposed; 
elastic effects are
taken into account in the framework of Maxwell model.  
The following three examples are studied using the proposed approach:
a transverse velocity autocorrelation function for 
free evolving system with random initial velocities,
a boundary-driven propagating shear waves, and a resonant
enhancement of oscillations in a periodically driven fluid in a 
capillary. The measured shear wave dispersion relation is found to be
in a good agreement with the theoretical one derived for the Navier-Stokes
equation with the Maxwell viscoelastic term. 
\end{abstract}
\begin{multicols}{2}

\section {Model}

Although only slightly more than a decade old, the Lattice Boltzmann (LB) 
method 
\cite{lb1,lb2,lb3,lbm}
has already gained the status
of a versatile simulation tool for homogeneous and heterogeneous flows 
in various, often very complex, geometries. 
Inclusion of viscoelastic effects, common
for many naturally-occurring fluids, will make the range of application
of the LB method even wider.
A  modification of the standard BGK model \cite {lb3}, suggested in 
\cite{ve}, allows for shear wave propagation, 
which is one of the intrinsic 
features of viscoelastic fluids. However, the physical foundations
of the approach, proposed in \cite{ve}, are somewhat unclear, since it does
not include any memory about an accumulated shear strain.
We propose a more general approach, based on a physically transparent 
Maxwell model of 
viscoelasticity \cite{m}, which exhibits viscoelastic properties and
accounts for accumulated stress via exponentially decaying memory function.

Let us for simplicity consider 
a standard 6-velocity BGK model
on a 2D hexagonal lattice, though generalizations for more sophisticated
lattice Boltzmann schemes is straightforward.
Evolution equations for an
{\it i}th channel occupation number $f_i$ have the following form \cite{lb3}:
\begin {equation}
\label{lb}
f_i(\vec{r}+\vec{c_i},t+1)=f_i(\vec{r},t)+\l \{ f_i(\vec{r},t)-
f_i^{eq}(\vec{r},t) \},
\end {equation}
where equilibrium occupation numbers $f_i^{eq}$ are:
\begin {equation}
f_i^{eq}={\r\over6}[1+2(\vec{C_i}\;\vec{U})+
G\ \{ (\vec{C_i}\;\vec{U})^2-{U^2\over2} \}].
\end {equation}
Here 
\begin {equation}
\r\equiv \sum_{i=1}^6 f_i,\; \;\vec{U}\equiv \sum_{i=1}^6 f_i \vec{C_i}
\end {equation}
are the equilibrium density and velocity at each lattice site, and  
$\vec{C_i}, \; i={1\ldots 6}$ are the lattice unit vectors.
Performing Chapman-Enskog expansion, one can prove \cite{lb3,lbm} 
that for $G=4$ 
these equations reproduce the Navier-Stokes equation
with correct convective term.

Maxwell model for viscoelastic media \cite{m}
links the elastic part of the 
stress tensor 
$\Pi_{ij}^{el}$ to the rate of strain $D_{ij}=\partial v_i/\partial x_j+
\partial v_j/\partial x_i $ via the linear equation
with exponentially decaying elastic ``memory'':
\begin {equation}
\label{maxwell}
{d \Pi_{ij}^{el}\over dt}= - {\Pi_{ij}^{el}\over \t}+{\m\over \t}D_{ij},
\end {equation}
where $\m$ is an elastic coefficient and $\t$ is a memory time.
As the 6-velocity BGK model adequately reproduces the Navier-Stokes equation
only for incompressible fluids $({\vec \nabla} {\vec v})=0$, 
in the following we limit our consideration to this case only.
Then the viscoelasticity of the fluid can be taken into account
by adding the
Maxwell elastic stress (body force)
term $\vec {F_el}(\vec{r},t)$, 
\begin {equation}
\label{Fnv}
\vec {F_{el}}(\vec{r},t)=\m \int_{-\infty}^t \exp[-(t-t')/\t]\;\,\Delta 
{\vec V} (\vec{r},t')
dt'/\,\t,
\end {equation}
to the right-hand side of Navier-Stokes equation.
In terms of Chapman-Enskog expansion \cite{lb3,lbm}
this elastic term has the same order ($\e^2$) 
as the standard viscous term. Hence to reproduce the elastic term
(\ref{Fnv}) in corresponding continuous (Navier-Stokes) equation we must add
its lattice equivalent to the relaxation term in the 
Lattice-Boltzmann equations (\ref{lb}):
\begin {eqnarray}
\label{upd}
\nonumber
f_i(\vec{r}+\vec{c_i},t+1)=f_i(\vec{r},t)+\l \{ f_i(\vec{r},t)-
f_i^{eq}(\vec{r},t) \}+\\
{1\over3}(\vec{F_{el}}(\vec{r},t)\;\vec{C_i}).
\end {eqnarray}
Here $\vec{F_{el}}$ is calculated similarly to (\ref{Fnv}), but
with the discretized time:
\begin {equation}
\label{F}
\vec{F_{el}}(\vec{r},t+1)=\vec{F_{el}}(\vec{r},t)[1-{1\over\t}]+\Delta \vec{U}(\vec{r},
t){\m\over\t},
\end {equation}
where $\Delta$ is the discrete Laplace operator
for the hexagonal lattice:
\begin {equation}
\label{du}
\Delta \vec{U}(\vec{r},t)={2\over3}\sum_{i=1}^6[\vec{U}(\vec{r}+\vec{C_i},t)-
\vec{U}(\vec{r},t)], 
\end {equation}
and $\vec{U}(\vec{r}_j)$ are equilibrium velocities at sites $\vec{r}_j$.

The Eqs.~(\ref{upd}-\ref{du}) formally define our model.
Thus, to include the Maxwell  
viscoelastic effects into the standard LB method,
one just needs to add a new local vector field (\ref{F}) to the 
standard BKG LB relaxation
term (\ref{upd}), which is updated each time step
for every lattice site $\vec{r}$. 
In the following section we first qualitatively and then
quantitatively check the proposed LB method 
(\ref{upd}) using three examples: a transverse velocity autocorrelation 
function for free evolving system with random initial velocities,
a boundary-driven propagating shear waves, and a resonant
enhancement of oscillations in a periodically driven fluid in a 
capillary.

\section {Simulations and dispersion relations}
To check whether our model have viscoelastic properties at all, 
we measure the Fourier transform of 
transverse velocity autocorrelation function
$|\tilde{U}_y(k_x,\om)|^2$, where
\begin {equation}
\label{trv}
\tilde{U}_y(k_x,\om) \equiv {1\over L^2 T}\sum_{x=-L/2}^{L/2}\sum_{y=1}^{L}
\sum_{t=1}^{T} U_y(x,y,t)e^{-ik_x x} e^{i\om t},
\end{equation}
where
\begin {equation} 
k_x=2\p n/L,\; n=-L/2, \ldots L/2,\;\; 
\om=2\p m/T,\;m=1,\ldots T.
\end {equation} 
Here the sum on $y$ corresponds to the averaging of $\tilde{U}_y(k_x,\om)$
on the $y$ coordinate of the sample. 
\begin{figure}
\centerline{\epsfxsize=8cm \epsfbox{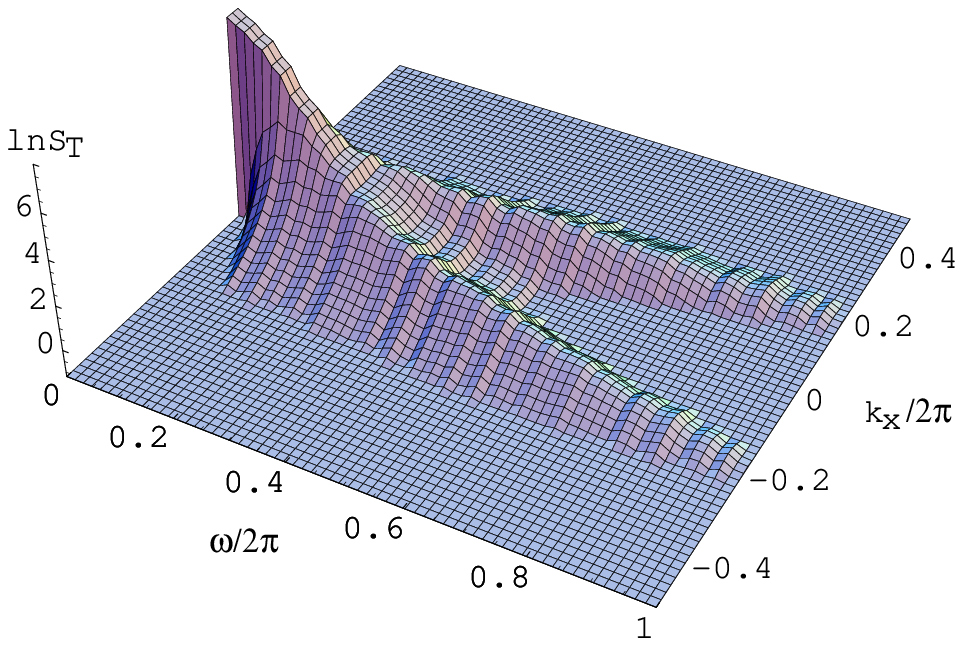}}
\noindent
{\small {\bf Fig.~1}. 
Sketch of the Fourier transform of 
the transverse velocity autocorrelation function,
$|\tilde{U}_y(k_x,\om)|^2$, for the LB system with elastic interaction.
The axes are $x=\om/2\pi$, $y=k_x/2\pi$, $z\propto \ln|\tilde{U}_y(k_x,\om)|$}
\end{figure}
\begin{figure}
\centerline{\epsfxsize=8cm \epsfbox{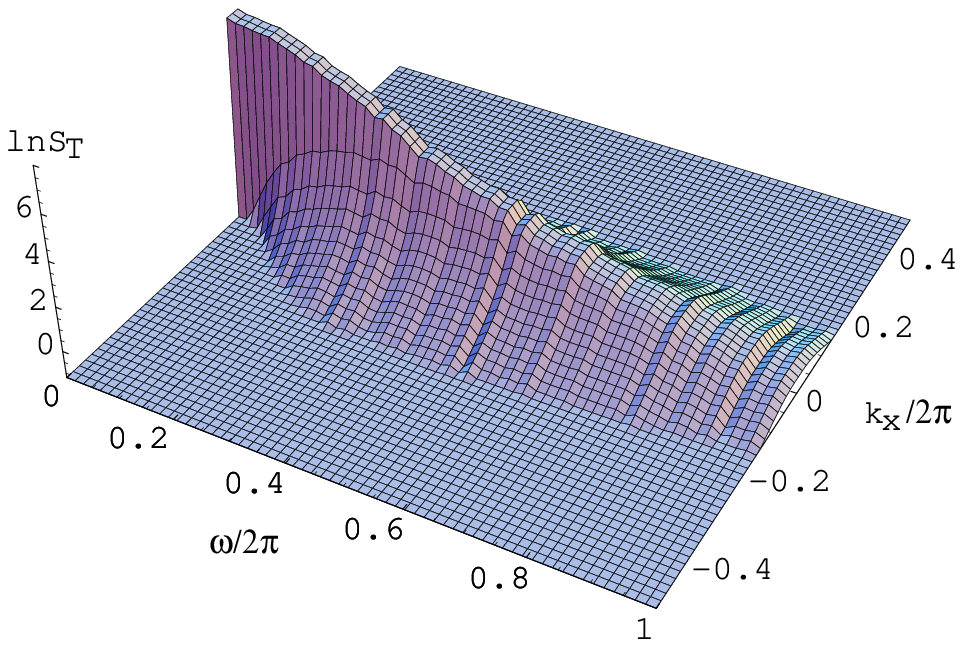}}
\noindent
{\small {\bf Fig.~2}.
Sketch of the Fourier transform of 
the transverse velocity autocorrelation function,
$|\tilde{U}_y(k_x,\om)|^2$, for the LB system without elastic interaction.
The axes are $x=\om/2\pi$, $y=k_x/2\pi$, $z=\propto \ln|\tilde{U}_y(k_x,\om)|$ }
\end{figure}
In our simulation we use the following parameters:
relaxation parameter $\l=-1.5$ (which corresponds to 
viscosity $\n=1/4(-1/\l-1/2)\approx 0.042$), elastic coefficient
$\m=0.3$, memory time $\t=10$, lattice of $256\times256$ sites, 
maximum time $T=256$ and random initial
occupation numbers corresponding to average
density per site $\r=1$. 
The simulation results, averaged over 
100 initial configurations, are presented in Fig.~1 and Fig.~2 for
the LB models with and without elastic effects, respectively.
Two symmetric branches $\om=\om(|k_x|)$, clearly noticeable in Fig.~1,
which correspond to propagating shear waves, indicate that the LB model
defined by (\ref{upd}) indeed exhibits viscoelastic properties.

To get more quantitative insight on the elastic feature of our model, we 
first
derive dispersion relations for the continuous 
shear waves in the framework of the 
Maxwell model.
Linearized Navier-Stokes equation with Maxwell elastic term 
for transverse velocity of incompressible fluid reads:
\begin {equation}
{\partial V \over \partial t}= \n \;\Delta V+
\m \int_{-\infty}^t \exp[-(t-t')/\t]\;\Delta V(t')
{dt'\over\t}.
\end {equation}
After inserting $V(x,t)=  V_0 \exp(-i\om t)\exp(ikx)$
and discarding exponentially decaying terms, we obtain
\begin {equation}
k^2=\om\; { i+\om t \over \n (1-i\om t)+\m}; 
\end {equation}
or separating real and imaginary parts of $k$, $\Re(k)$ and $\Im(k)$:
\begin {eqnarray}
\label{ri}
\nonumber
k=\sqrt{{\om \over2[(\m+\n)^2+(\n \om\t)^2]}}\times\\
\left\{
\sqrt{\m\om\t+\sqrt{(\m\om\t)^2+[\n(1+\om^2\t^2)+\m]^2}}+
\right. \\
\nonumber
\left.
i\sqrt{-\m\om\t+\sqrt{(\m\om\t)^2+[\n(1+\om^2\t^2)+\m]^2}}
\right\}
\end {eqnarray}

We are interested in propagating shear waves with sufficiently small 
dissipation, 
hence we consider parameters for which the ratio
\begin {equation}
{\Im(k)\over\Re(k)}={\n[1+(\om\t)^2]+\m \over
\m\om\t+\sqrt{(\m\om\t)^2+[\n(1+\om^2\t^2)+\m]^2}}
\end{equation}
is small.
Naively, one can set $\n\om^2\t^2 \ll \m$ and
$\om\t\rightarrow +\infty$, thus setting 
$\Im(k)/\Re(k)\rightarrow (2\om\t)^{-1}\rightarrow 0$.
But LB systems with too small viscosity 
and too large elastic 
coefficient turn out to be numerically unstable. Either for random initial 
conditions
or for small drive, we experimentally determined that domain of stability is 
roughly limited by $\n\geq 0.04$ and $\m\leq 0.3$ 
For these boundary values of viscosity and elastic coefficient we find that
${\Im{(k)}/\Re{(k)}}$ reaches its minimum when $\om\t=\phi \approx 2.9$, 
at which 
\begin{equation}
\label{pred}
\Re{(k)}\approx 0.69, \;\; \Im{(k)}\approx 0.24.  
\end{equation}
To check whether our model reproduces this
continuous prediction, we perform the following simulation.
We consider a $100\times 100$ lattice 
with 
periodic boundary conditions in $X$ direction and reflecting 
boundary conditions  in $Y$ direction.
The reflection form the $y=0$ wall is periodically modulated so that 
the $X$ components of the velocities after reflection were set to be 
proportional to
$\cos{(\om t)}$. Simulations were performed for $\l=-1.5$, (or 
$\n=1/4[-1/\l-1/2]\approx 0.042$), elastic coefficient
$\m=0.3$, memory time $\t=46$, and the period of oscillation $T$ 
corresponded to
the theoretical minimum of ${\Im{(k)}/\Re{(k)}}$,
$T\equiv 2\p\t/\phi\approx 100$. 
\begin{figure}
\centerline{\epsfxsize=8cm \epsfbox{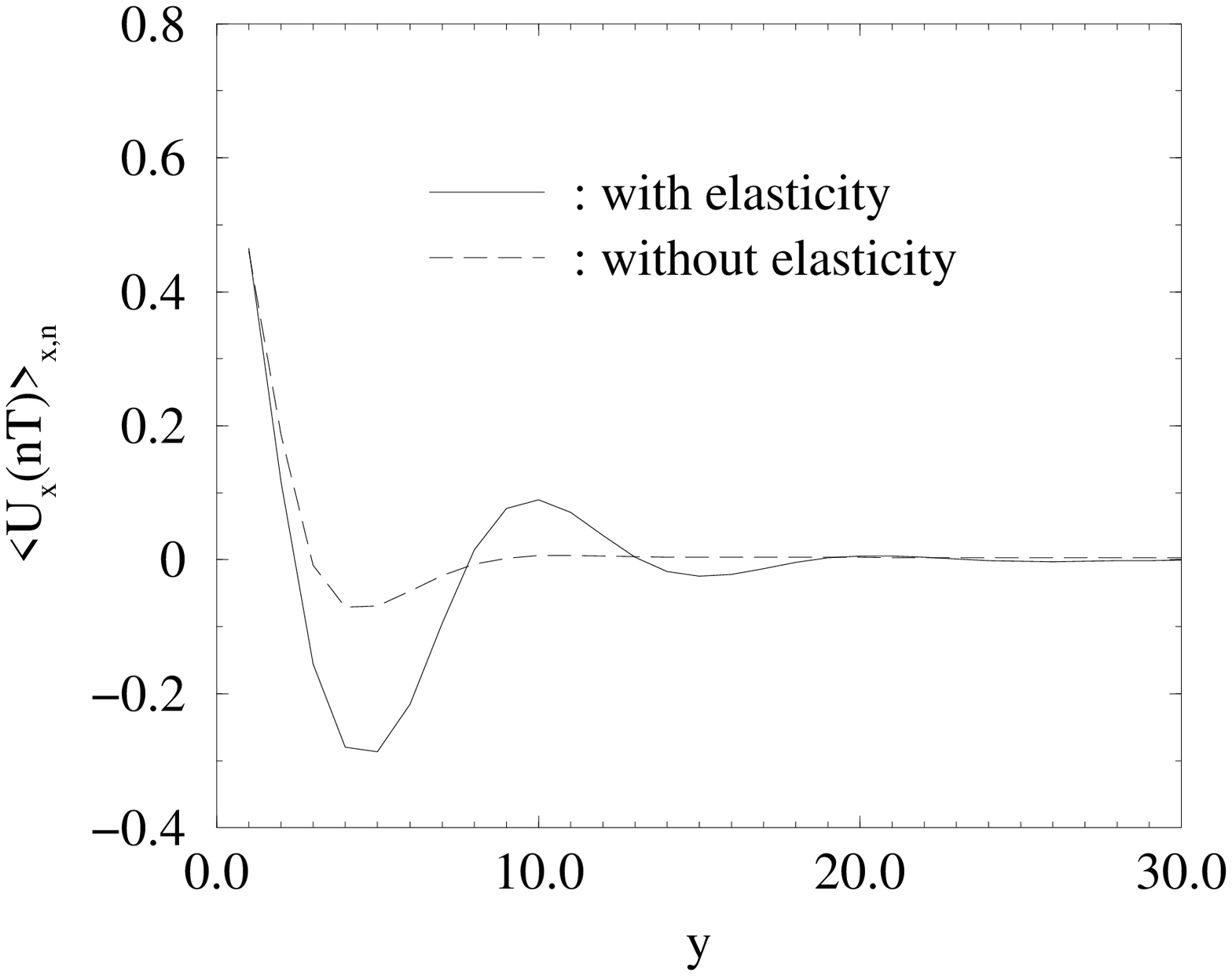}}
\noindent
{\small {\bf Fig.~3}. 
Plot of the $X$ component of the average velocity 
$\langle V_x(y,t_n)\rangle_{x,n}$ measured at the times
$t_n=nT$, $n=1,\ldots$ vs. the distance form the driving wall
$y$.}
\end{figure}
\begin{figure}
\centerline{\epsfxsize=8cm \epsfbox{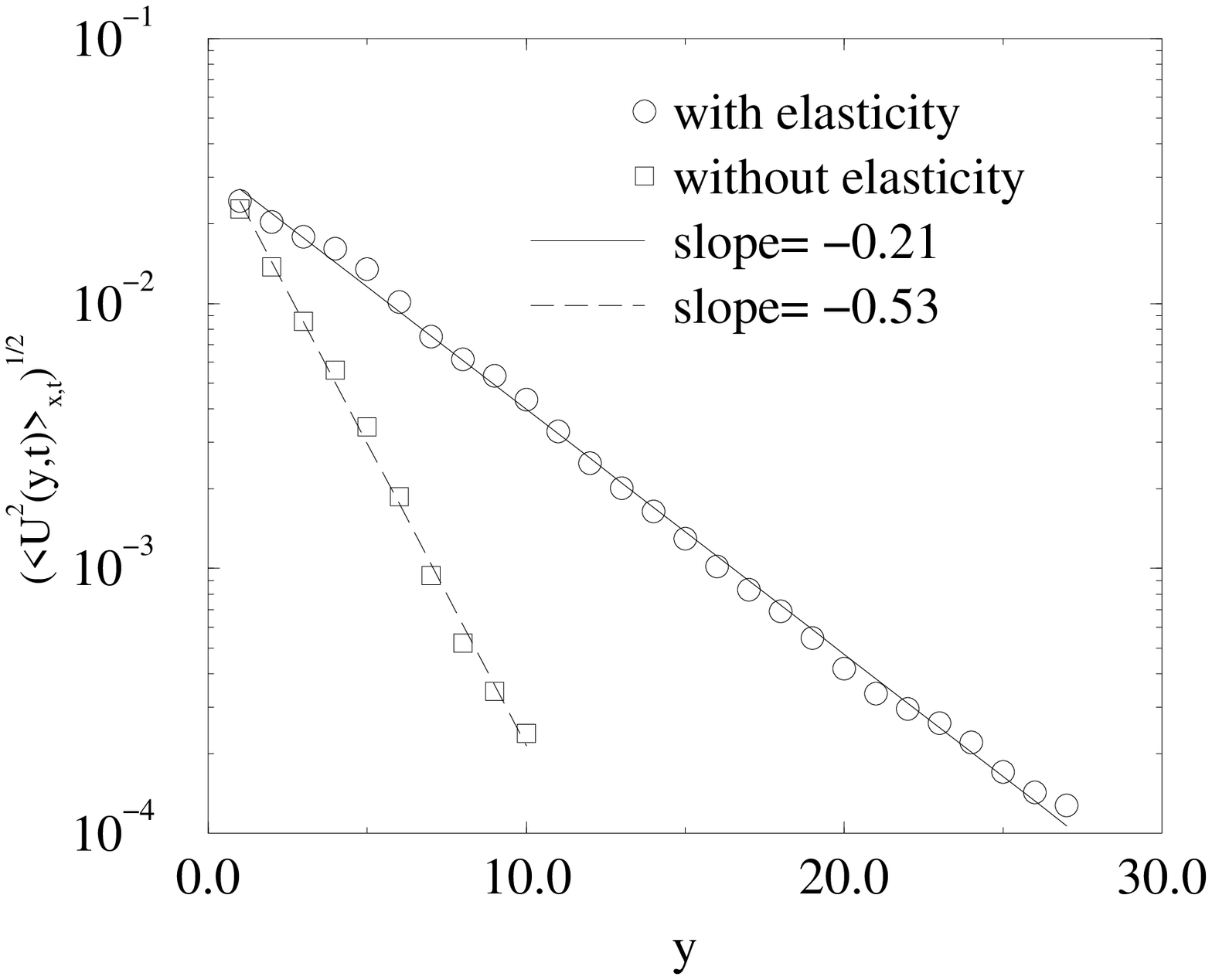}}
\noindent
{\small {\bf Fig.~4}. 
Square root of the mean square
velocity $\sqrt{\langle V_x^2(y,t)\rangle_{x,t}}$ 
as a function of distance from the driving wall $y$.}
\end{figure}
The simulation results are presented in the figures 3 and 4.
Fig.~3 shows the plot of the $x$ component of velocity 
$\langle V_x(y,t_n)\rangle_{x,n}$, measured at the times
$t_n=nT$, $n=1,\ldots$ and averaged over $x$ and $n$. 
In the Fig.4 the square root of mean square
velocity
$\sqrt{\langle V_x^2(y,t)\rangle_{x,t}}$ is plotted as a 
function of distance from the wall $y$.
We observe that $\Re{(k)}\approx 0.63$ (wavelength of oscillation is equal to
10) and $\Im{(k)}\approx0.21$, which, given the discrete nature of the LB
simulations, is in good agreement with the theoretical values (\ref{pred})
For illustrative purposes, Figs ~3,4 contain results obtained using similar
system, but without the elastic
effects ($\m=0$), for which ${\Im{(k)}/\Re{(k)}=1}$.

Finally, 
let us consider one more example on  which the viscoelastic properties 
of our LB approach
can be quantitatively studied:
a periodically volume-driven fluid in a capillary. The model consists
of a $9\times 128$ elongated lattice with stick boundary conditions for 
long walls and periodic boundary conditions for short walls. A uniform 
time-periodic 
volume force $F_y=F_0 \cos (\om t)$, directed along the longer walls, is 
applied to the fluid in the capillary. 
In the LB update scheme (\ref{upd}) it is implemented by adding 
the driving force $F_y(t)$ to the elastic force $\vec F_{el}$
in the relaxation term on the right hand side of the lattice equation. 

As the fluid in the capillary exhibits shear elasticity, there should be a 
resonance when the driving frequency $\om $ coincides with the intrinsic
oscillation frequency $\Omega$ of elastic media in the capillary.
To study the resonance, we look at the driven oscillations of the first 
Fourier harmonic
of velocity along the applied force,
\begin {equation}
\tilde V_{1y}(t)={2\over L_x L_y}\int_0^{L_x}\int_0^{L_y}\sin({x\over L_x}\p)
V_y(x,y,t)dxdy,
\end {equation}
where the integration on $dy$ stands for averaging along the length
of the capillary.
In the presence of volume force $F_y(t)$, a 
linearized Navier-Stokes equation yields for
the steady state amplitude of the first Fourier harmonic $\tilde V_{1y}$:
\begin {equation}
\label{amp}
\tilde V_{1y}={4F_0\over q L_x}{1\over q^2 [\n+{\m\over 1+\om^2 \t^2}] 
+i\om[{q^2\m\t\over 1 + \om^2 \t^2}-1]},
\end {equation}
where $q=\pi/L_x$ is a wavevector of the first harmonic.
The resonance is achieved when the absolute value of this expression 
has the maximum, i.e.
when the second bracket in denominator becomes equal to zero, 
\begin {equation}
\label{res}
\om=\Omega\equiv {1\over\t}\sqrt{q^2\m\t-1}
\end {equation}
Also, when a system is in 
resonance, there no phase shift between driving force and induced 
oscillation so that the ratio between the force and the oscillation 
amplitudes is real. 
To check whether our LB model behaves as predicted by the continuous
Navier-Stokes equation, 
we performed a simulation using the following parameters:
$\m=0.3$, $\l=-1.5$, $\t=250$, $F_0=0.001$; frequency of the drive $\om$ was 
equal to $\Omega\approx 0.011$, $\Omega/2$, and $2\Omega$.
The measured values of the $\tilde V_{1y}$ are presented in Fig.5.
\begin{figure}
\centerline{\epsfxsize=8cm \epsfbox{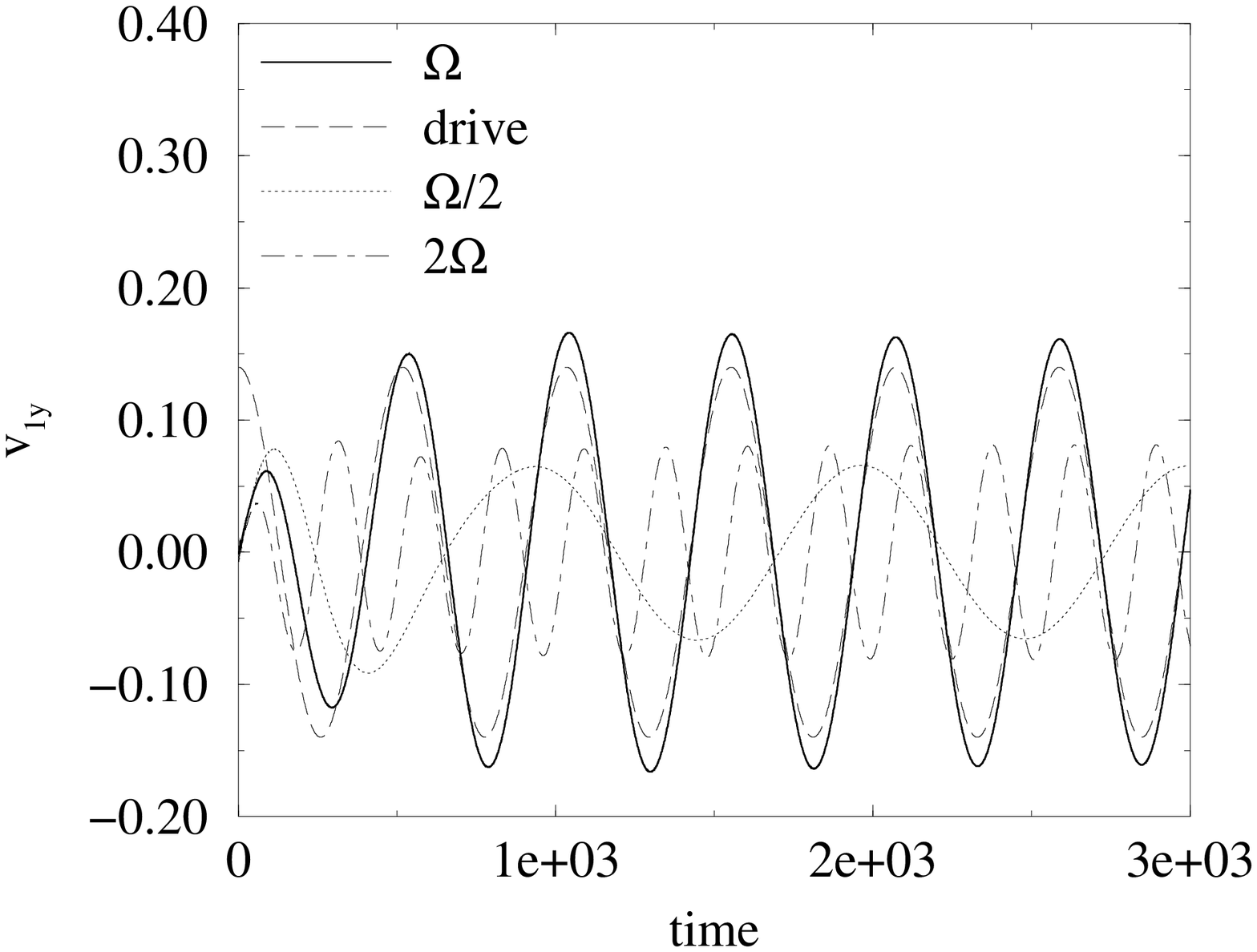}}
\noindent
{\small {\bf Fig.~5}. Amplitude of the first Fourier harmonic
$\tilde V_{1y}$ as a function of time for different driving frequencies,
$\om=\Omega$, $\om=2\Omega$, $\om=\Omega/2$.}
\end{figure}
After a short transient 
period the system approaches the steady state. As predicted by (\ref{amp}),  
for resonant drive ($\om=\Omega$) the steady state amplitude of oscillation
$\tilde V_{1y}\approx 0.16$; while for off-resonance drive,
 $\om=\Omega/2$ and $\om=2 \Omega$, the  amplitudes 
$\tilde V_{1y}$ are at least twice smaller.
To study a phase shift between the drive and induced oscillations, 
we also plotted a driving force for the resonant frequency,
$F(t)\sim \cos(\Omega t)$. One can observe that there is no visible phase
shift between driving and induced oscillation, which confirms that
the shear elastic resonance frequency of our LB model is
well reproduced by the continuous expression (\ref{res}).
We also observe that when the drive frequency $\om$ is below or above the 
resonance frequency $\Omega$, the induced oscillations are
phase-delayed and  phase-advanced, respectively, which again follows
from  (\ref{res}).

Three examples, considered in this section, indicate that for a variety
of physical systems, the behavior of the proposed viscoelastic LB model is
qualitatively and quantitatively described by the continuous
Navier-Stokes equations with Maxwell viscoelastic term. It allows us to
conclude that our model indeed reproduces the
Maxwell viscoelasticity for the incompressible fluid flow.

\section {Conclusion}

We proposed a simple yet versatile approach of incorporating
viscoelastic effects into Lattice Boltzmann simulations for 6-velocity
2D BGK model
Generalization for more sophisticated LB schemes which includes
3D and higher-velocity models is straightforward.
Although, for computational stability and efficiency reasons,
our approach has not yet allowed us to reach a limit of pure 
elastic solids, it works well in the range of parameters typical for most 
naturally-occurring viscoelastic media.
We leave consideration of more sophisticated than the Maxwell
viscoelastic models for the future.

\end{multicols}
\end{document}